\documentclass{iau_FM}
\usepackage{graphicx}
\usepackage{framed,color,fancy box,wrap fig,scalefnt,multicol}

\title 
{Resolved Host Studies of Stellar Explosions}

\author[Emily M. Levesque]   
{Emily M. Levesque$^1$}

\affiliation{$^1$University of Washington Department of Astronomy, Box 351580, Seattle, WA 98195 USA \\ email: {\tt emsque@uw.edu} }
\pubyear{2015}
\jname{Astronomy in Focus, Volume 1} 
\begin{document}

\maketitle

\begin{abstract}
The host galaxies of nearby (z$<$0.3) core-collapse supernovae and long-duration gamma-ray bursts offer an excellent means of probing the environments and populations that produce these events' varied massive progenitors. These same young stellar progenitors make LGRBs and SNe valuable and potentially powerful tracers of star formation, metallicity, the IMF, and the end phases of stellar evolution. However, properly utilizing these progenitors as tools requires a thorough understanding of their formation and, consequently, the physical properties of their parent host environments. In this talk I will review some of the recent work on LGRB and SN hosts with resolved environments that allows us to probe the precise explosion sites and surrounding environments of these events in incredible detail.
\end{abstract}

The luminous core-collapse deaths of massive stars as long-duration gamma-ray bursts (LGRBs) can serve as valuable probes of their star-forming galaxies stretching out to high redshifts. However, for the most nearby events these host galaxies are also the best windows we have for getting a glimpse at the nature of these events' progenitors. At distances where pre-explosion imaging, light echo, and supernova remnants studies are impossible (with current facilities), the hosts of LGRBs allow us to probe the metallicities, ages, and formative environments of their parent stellar populations.

For most LGRBs we are limited to global studies of the host galaxies as a whole, examining morphological trends in LGRB localization (e.g. Fruchter et al.\ 2006) or obtaining single spectra that represent a galaxy-wide composite of interstellar medium properties (Levesque et al.\ 2010a). However, a small sample of LGRBs have occurred in galaxies that are either nearby enough or massive enough to facilitate resolved comparisons between the precise stellar explosion site and other regions of the host galaxy. Currently, six of these events have had their hosts studied in detail.

GRBs 980425, 100316D, and 120422A are all ``subluminous" GRBs, with much lower energies and luminosities than the ``cosmological" GRB population (e.g. Stanek et al.\ 2006). All three occurred at $z < 0.3$ and were accompanied by well-studied Type Ic-BL SNe. Studies of all three hosts demonstrated that the galaxies were metal-poor with relatively weak metallicity gradients; while the GRB explosion sites had slightly lower metallicities that the rest of the galaxies, the differences were within the errors of the metallicity diagnostics (e.g. Christensen et al. 2008, Levesque et al. 2011, Levesque et al. 2012). The GRBs were also localized near the strongest star-forming regions of their hosts, showing evidence of very young massive star populations (e.g. Levesque et al.\ 2011, Le Floc'h et al.\ 2012, Michalowski et al.\ 2014).

GRB 060505 was another subluminous GRB at $z\sim0.089$; however, it lacked {\it any} detected signature of an accompanying supernova and also showed an unusually short burst duration of only 4s (Fynbo et al.\ 2006). Despite this, spatially resolved spectra showed that, like the GRB/SNe described above, GRB 060505 was localized in a star-forming low-metallicity galaxy, in an explosion site showing signs of recent strong star formation and with a metallicity comparable to that of the host (Th\"one et al. 2008).

By contrast, GRB 020819 had a ``cosmological" luminosity more typical of the general LGRB population and was localized in a much more distant and massive star-forming spiral galaxy, at $z\sim0.41$. It was also classified as a ``dark" GRB, which are characterized by drop-outs in the optical regimes of their afterglows (e.g. Perley et al.\ 2013); however, radio afterglow observations were able to localize the explosion site of this burst in a small star-forming region located in one of the galaxy's spiral arms. Observations showed that both the host nucleus and the explosion site had comparable metallicities; however, both were also quite high, marking this as the first confirmed observation of an LGRB produced in a high-metallicity environment (Levesque et al.\ 2010b).\begin{wrapfigure}{L}{0.4\textwidth}
\vspace{-10pt}
\includegraphics[width=0.4\textwidth]{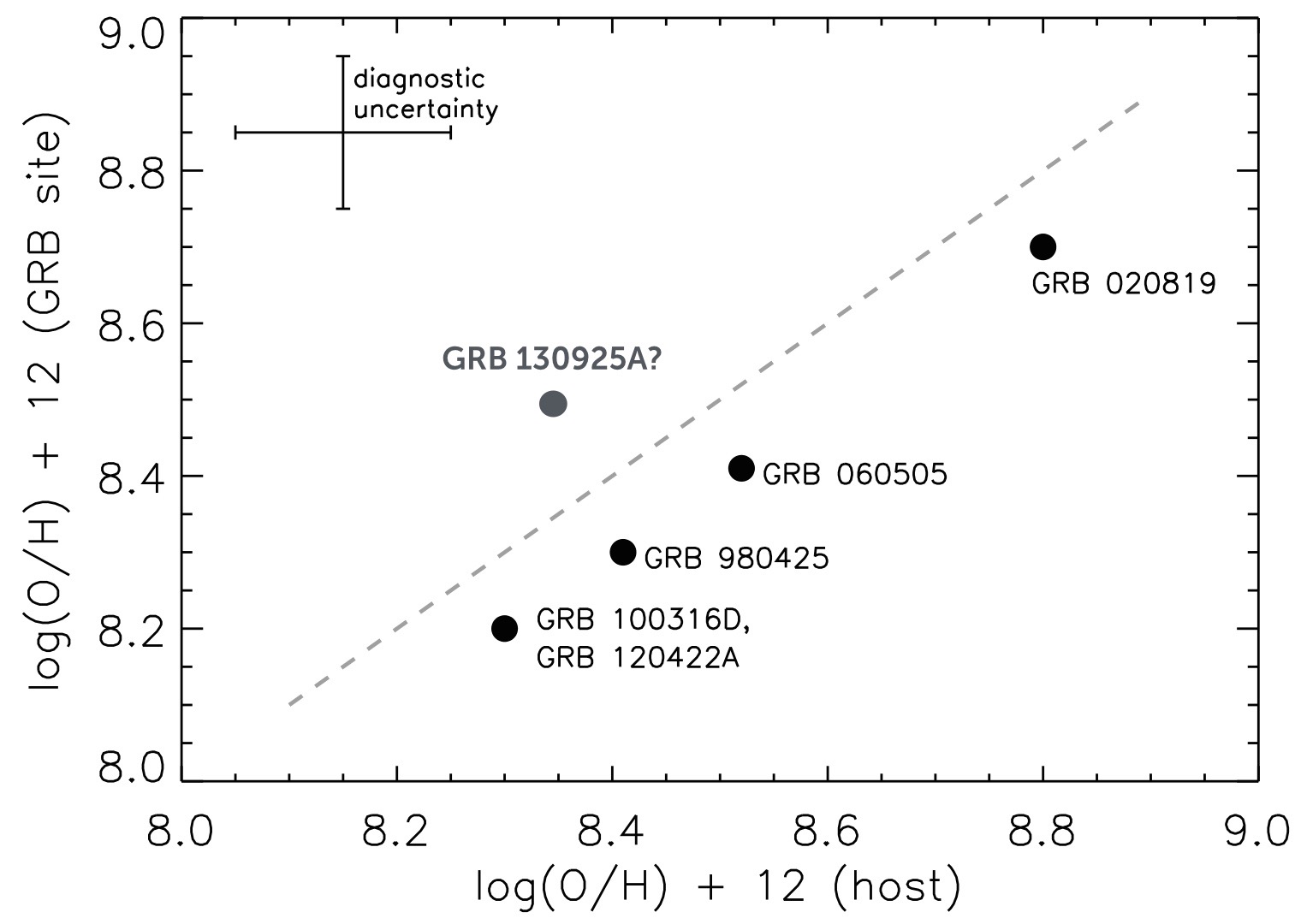}
\vspace{-10pt}
{\it \scalefont{0.5} \caption{Comparison of host and explosion site metallicities for LGRBs as determined from Pettini \& Pagel (2004) diagnostics. The relation where site and host metallicities are identical is plotted as a gray dashed line.}}
\vspace{-10pt}
\end{wrapfigure}

Finally, GRB 130925A is a member of the recently-identified class of ultra-long GRBs (ULGRBs). Only four examples of ULGRBs have been identified to date , all with durations exceeding 1000s; with a duration of $\sim$20,000s, GRB 130925A is the longest of this sample. It is still unclear whether ULGRBs represent a phenomenologically distinct class of transient or one extreme in a continuum of high-energy events. The host of GRB 130925A is very dusty and metal-rich; however, these properties are similar to those seen in typical LGRB host galaxies (e.g. Levesque et al.\ 2010b, Perley et al.\ 2013), and the explosion site metallicity is once again comparable to other star-forming regions seen within the host (Schady et al.\ 2015).

Figure 1 shows a comparison of these six LGRBs' explosion site and host metallicities. All six events' explosion sites show agreement with the theoretical relation where explosion site and host metallicities are identical to within the errors of the diagnostics. From these results it appears that global host metallicities for star-forming galaxies are indeed representative of LGRB explosion site metallicities, an encouraging result considering that only global host observations are possible at higher redshifts. However, the small sample precludes drawing any definitive conclusions, and further studies of explosion sites and metallicity gradients in LGRB host galaxies are needed.

There are currently six remaining galaxies where such detailed studies are possible: the hosts of GRBs 990705, 011121, 020903, 030329, 060218, and 130425A (while other nearby bursts exist, the hosts are too compact for spatially resolved analyses). Although the overall sample remains small, resolved studies of these galaxies would double the current sample. This in turn would allow us to further explore the metallicity and star formation properties of LGRB explosion sites and host galaxies as a whole, clarifying our understanding of the interstellar medium conditions that give rise to LGRB progenitors and how these events can be utilized as probes of star-forming galaxies at high redshift.

\begin{multicols}{2}
\noindent {\bf References} \\
Christensen, L., et al.\ 2008, A\&A, 490, 45 \\
Fruchter, A.S., et al.\ 2006, Nature, 441, 463 \\
Fynbo, J.P.U., et al.\ 2006, Nature, 441, 465 \\
Le Floc'h, E., et al.\ 2012, ApJ, 746, 7 \\
Levesque, E. M., et al.\ 2010a, AJ, 140 1557 \\
Levesque, E. M., et al.\ 2010b, ApJ, 712, L26 \\
Levesque, E. M., et al.\ 2011, ApJ, 739, 23 \\
Levesque, E. M., et al.\ 2012, ApJ, 758, 92 \\
Michalowski, M. J., et al.\ 2014, A\&A, 571, 75 \\
Pettini, M. \& Pagel, B. E. J. 2004, MNRAS, 348, 59 \\
Perley, D. A., et al.\ 2013, ApJ, 778, 128 \\
Schady, P., et al.\ 2015, A\&A, 579, 126\\
Stanek, K. Z., et al.\ 2006, Acta Astron., 56, 333 \\
Th\"{o}ne, C. C., et al.\ 2008, ApJ, 676, 1151 
\end{multicols}

\end{document}